\documentclass[]{pasj01}

\begin{document} 
\Received{}
\Accepted{}

\title{J-GEM Follow-Up Observations of The Gravitational Wave Source
GW151226\altaffilmark{1}}

\altaffiltext{1}{Based on data collected at the Subaru Telescope, which is operated by the
National Astronomical Observatory of Japan.}

\author{Michitoshi \textsc{YOSHIDA}\altaffilmark{2}}
\altaffiltext{2}{Hiroshima Astrophysical Science Center, Hiroshima University, Hiroshima 739-8526, Japan}
\email{yoshidam@hiroshima-u.ac.jp}

\author{Yousuke \textsc{UTSUMI},\altaffilmark{2}}

\author{Nozomu \textsc{TOMINAGA}\altaffilmark{3,4}}
\altaffiltext{3}{Department of Physics, Faculty of Science and Engineering, Konan University, Kobe, Hygo 658-8501, Japan}
\altaffiltext{4}{Kavli Institute for the Physics and Mathematics of the Universe (WPI), The University of Tokyo, Kashiwa, Chiba 277-8583, Japan}

\author{Tomoki \textsc{MOROKUMA}\altaffilmark{5,4}}
\altaffiltext{5}{Institute of Astronomy, Graduate School of Science, The University of Tokyo, Mitaka, Tokyo 181-0015, Japan}

\author{Masaomi \textsc{TANAKA}\altaffilmark{6,4}}
\altaffiltext{6}{National Astronomical Observatory of Japan, Mitaka, Tokyo 181-8588, Japan}

\author{Yuichiro \textsc{ASAKURA}\altaffilmark{7}}
\altaffiltext{7}{Institute for Space-Earth Environmantal Research, Nagoya University, Chikusa-ku, Nagoya 464-8601, Japan}

\author{Kazuya \textsc{MATSUBAYASHI}\altaffilmark{8}}
\altaffiltext{8}{Department of Astronomy, Kyoto University, Kitashirakawa-Oiwake Kyoto 606-8502, Japan}

\author{Kouji \textsc{OHTA}\altaffilmark{8}}

\author{Fumio \textsc{ABE}\altaffilmark{7}}

\author{Sho \textsc{CHIMASU}\altaffilmark{9}}
\altaffiltext{9}{Department of Physics, School of Science, Tokai University, Hiratsuka, Kanagawa 259-1292, Japan}

\author{Hisanori \textsc{FURUSAWA}\altaffilmark{6}}

\author{Ryosuke \textsc{ITOH}\altaffilmark{10,11}}
\altaffiltext{10}{Department of Physical Science, Hiroshima University, Hiroshima 739-8526, Japan}
\altaffiltext{11}{Department of Physics, Tokyo Institute of Technology, Meguro-ku, Tokyo 152-8551, Japan}

\author{Yoichi \textsc{ITOH}\altaffilmark{12}}
\altaffiltext{12}{Nishi-Harima Astronomical Observatory, University of Hyogo, Hyogo 679-5313, Japan}

\author{Yuka \textsc{KANDA}\altaffilmark{10}}

\author{Koji S. \textsc{KAWABATA}\altaffilmark{2}}

\author{Miho \textsc{KAWABATA}\altaffilmark{10}}

\author{Shintaro \textsc{KOSHIDA}\altaffilmark{13}}
\altaffiltext{13}{Subaru Telescope, National Astronomical Observatory of Japan, 650 North Afohoku Place, Hilo, HI 96720, USA}

\author{Naoki \textsc{KOSHIMOTO}\altaffilmark{14}}
\altaffiltext{14}{Department of Earth and Space Science, Graduate School of Science, 
Osaka University, Toyonaka, Osaka 560-0043, Japan}

\author{Daisuke \textsc{KURODA}\altaffilmark{15}}
\altaffiltext{15}{Okayama Astrophysical Observatory, National Astronomical Observatory of Japan, Asakuchi, Okayama 719-0232, Japan}

\author{Yuki \textsc{MORITANI}\altaffilmark{4}}

\author{Kentaro \textsc{MOTOHARA}\altaffilmark{5}}

\author{Katsuhiro L. \textsc{MURATA}\altaffilmark{16}}
\altaffiltext{16}{Department of Particle and Astrophysical Science, Nagoya University, Chikusa-ku, Nagoya 464-8602, Japan}

\author{Takahiro \textsc{NAGAYAMA}\altaffilmark{17}}
\altaffiltext{17}{Graduate School of Science and Engineering, Kagoshima University, Kogoshima 890-0065, Japan}

\author{Tatsuya \textsc{NAKAOKA}\altaffilmark{10}}

\author{Fumiaki \textsc{NAKATA}\altaffilmark{13}}

\author{Tsubasa \textsc{NISHIOKA}\altaffilmark{18}}
\altaffiltext{18}{Department of Physics, Faculty of Science, Kyoto Sangyo University, 603-8555 Kyoto, Japan}

\author{Yoshihiko \textsc{SAITO}\altaffilmark{11}}

\author{Tsuyoshi \textsc{TERAI}\altaffilmark{13}}

\author{Paul J. \textsc{TRISTRAM}\altaffilmark{19}}
\altaffiltext{19}{Mt. John University Observatory, Lake Tekapo 8770, New Zealand}

\author{Kenshi \textsc{YANAGISAWA}\altaffilmark{15}}

\author{Naoki \textsc{YASUDA}\altaffilmark{4}}

\author{Mamoru \textsc{DOI}\altaffilmark{5,20}}
\altaffiltext{20}{Research Center for the Early Universe, Graduate School of Science, 
The University of Tokyo, Bunkyo-ku, Tokyo 113-0033, Japan}

\author{Kenta \textsc{FUJISAWA}\altaffilmark{21}}
\altaffiltext{21}{The Reseach Institute of Time Studies, Yamaguchi University, Yamaguchi 753-8511, Japan}

\author{Akiko \textsc{KAWACHI}\altaffilmark{9}}

\author{Nobuyuki \textsc{KAWAI}\altaffilmark{11}}

\author{Yoichi \textsc{TAMURA}\altaffilmark{5}}

\author{Makoto \textsc{UEMURA}\altaffilmark{2}}

\author{Yoichi \textsc{YATSU}\altaffilmark{11}}

\KeyWords{gravitational waves --- black hole physics --- surveys --- methods:observational --- binaries:close} 

\maketitle

\begin{abstract}
We report the results of optical--infrared follow-up observations of the gravitational wave (GW) 
event GW151226 detected by the Advanced LIGO in the framework of 
J-GEM (Japanese collaboration for Gravitational wave ElectroMagnetic follow-up). 
We performed wide-field optical imaging surveys with Kiso Wide Field Camera (KWFC),
Hyper Suprime-Cam (HSC), and MOA-cam3.
The KWFC survey started at 2.26 days after the GW event and covered 778 deg$^2$
centered at the high Galactic region of the skymap of GW151226. 
We started the HSC follow-up observations from $\sim$12 days after the event and
covered an area of 63.5 deg$^2$ of the highest probability region of the northern sky
with the limiting magnitudes of 24.6 and 23.8 for $i$ band and $z$ band, respectively.
MOA-cam3 covered 145 deg$^2$ of the skymap with MOA-red filter 
$\sim$2.5 months after the GW alert.
Total area covered by the wide-field surveys was 986.5 deg$^2$.
The integrated detection probability of all the observed area was $\sim$29\%.
We also performed galaxy--targeted observations with six optical and near-infrared
telescopes from 1.61 days after the event.
Total of 238 nearby ($\leq$100 Mpc) galaxies were observed 
with the typical $I$ band limiting magnitude of $\sim$19.5.
We detected 13 supernova candidates with the KWFC survey, and 60 extragalactic
transients with the HSC survey.
Two third of the HSC transients were likely supernovae 
and the remaining one third were possible active galactic nuclei.
With our observational campaign, we found no transients
that are likely to be associated with GW151226.
\end{abstract}

\section{Introduction}

Gravitational wave (GW) is a quadrupole wave of space-time distortion propagating with the light speed. 
Strong GW is emitted by violent gravitational disturbance induced by a coalescence
between 
compact massive objects such as neutron stars (NSs) or black holes (BHs).
In order to observe GW directly,
new generation GW detectors; Advanced LIGO (aLIGO, \cite{Abbott2016b}), 
Advanced Virgo (aVirgo; \cite{Acernese2015}), and KAGRA \citep{Somiya2012}
are being constructed.
If the planned sensitivities are achieved, these GW detectors can detect GW signals from an 
NS--NS merger at a distance of 200 Mpc \citep{Abadie2010}.
The GW detection rate is anticipated to be in a range of 0.4--400 events yr$^{-1}$ for 
NS--NS merger \citep{Abadie2010}.
Uncertainty of the above number primarily comes from 
the limit of our knowledge on real number of NS binary fraction in a galaxy.

If a compact object merger contains one NS, wide wavelength range of electromagnetic (EM)
emission associated with GW is expected \citep{Li1998,Rosswog2005,Nakar2011,Roberts2011,Metzger2010,
Metzger2012,Barnes2013,Hotokezasa2013,Tanaka2013,Berger2014,Tanaka2014}.
The EM emission would tell us important pieces of information about the nature of the GW event;
its astrophysical origin, detailed localization, accurate distance, and local environment of the event.
Most promising optical--near-infrared emission from GW sources is radioactively-powered emission, 
so called ``kilonova'' or ``macronova'' 
associated with NS--NS or BH--NS mergers \citep{Metzger2012,Barnes2013,Tanaka2014}.
A strong tidal force induced by merging process blows out the outer layer of NS,
and a wide solid angle outflow from the merger emits a wide range of EM emission due to 
radioactive decay of the ejecta, that is ``kilonova''. 
Neutron rich ejecta of a kilonova produce huge amount of r-process elements,
thus the kilonova emission gives important clues to the long standing mystery about the 
sites of cosmic r-process nucleosynthesis.
Moreover, the luminosity and light curve of a kilonova would allow us to constraint 
the equation of state of NS.
To search for EM emission associated with GW events,
we organized an EM follow-up observation network J-GEM (Japanese collaboration of Gravitational
wave Electro-Magnetic follow-up; \cite{Morokuma2016}) by utilizing optical-infrared-radio 
telescopes of Japan.

The first direct detection of GW was achieved by aLIGO on Sep. 14 2015 \citep{Abbott2016a}.
aLIGO performed the first science run (O1) from Sep. 2015 to Jan. 2016.
Just before the regular operation of O1, aLIGO detected the GW at 
Sep. 14 2015 09:50:45 UT \citep{Abbott2016a}.
The GW from this event, which was named as GW150914, was emitted by a 36 $M_\odot$--29 $M_\odot$ binary BH coalescence.  
While many electromagnetic (EM) follow-up observations were performed for GW150914
\citep{Abbott2016d,Abbott2016e,Ackermann2016,Evans2016a,Kasliwal2016,Lipunov2016,
Morokuma2016,Serino2016,Smartt2016a,
Sores-Santos2016,Troja2016},
no clear EM counterpart was identified with those observations except for a possible
detection of $\gamma$-ray emission by $Fermi$ Gamma-ray Burst Monitor (GBM) \citep{Connaughton2016}.
However, the $Fermi$ GBM detection was not confirmed by INTEGRAL observations \citep{Savchenko2016}.

aLIGO detected another GW signal during O1.
This event was detected at 03:38:53 UT on Dec. 26 2015 and was named as GW151226.
The false alarm probability of the event was estimated as $<$10$^{-7}$ ($>$5$\sigma$) and 
3.5$\times$10$^{-6}$ (4.5$\sigma$) \citep{Abbott2016c}.
The GW was also attributed to a BH--BH binary merger whose masses are 14.2$^{+8.3}_{-3.7}$ $M_\odot$ 
and 7.5$^{+2.3}_{-2.3}$ $M_\odot$.
The final BH mass was 20.8$^{+6.1}_{-1.7}$ $M_\odot$ and a gravitational energy of 
$\sim$1 $M_\odot$ was emitted as GW.
The distance to the event was 440$^{+180}_{-190}$ Mpc \citep{Abbott2016c}.

Here, we report the EM counterpart search for GW151226 performed in the framework of J-GEM.
We assume that cosmological parameters $h_0$, $\Omega_m$, and $\Omega_\lambda$
are 0.705, 0.27, and 0.73, respectively \citep{Komatsu2011} in this paper.
All the photometric magnitudes presented in this paper are AB magnitudes. 


\section{Observations}

We performed wide-field survey and galaxy targeted follow-up observations in and around the probability 
skymap of GW151226.
The 90\%\ credible area of the initial skymap created by BAYESTAR algorithm \citep{Singer2014}
was $\sim$1400 deg$^2$ \citep{LSC2015}.
The final skymap was refined by LALInference algorithm \citep{Veitch2015} and
the 90\%\ area is finally 850 deg$^2$ \citep{Abbott2016c}.
We also made an integral field spectroscopy for an optical transient (OT) candidate reported by
MASTER.
The specifications of the instruments and telescopes we used for the follow-up observations
are summarized in \citet{Morokuma2016}.

\subsection{Wide Field Survey}

We used three instruments for the wide-field survey; KWFC \citep{Sako2012} on the
1.05 m Schmidt telescope at Kiso Observatory, HSC \citep{Miyazaki2012} on the 8.2 m
Subaru Telescope, and MOA-cam3 \citep{Sako2008} on
the 1.8 m MOA-II telescope at Mt. John Observatory
in New Zealand.

\begin{figure}
 \begin{center}
  \includegraphics[width=8cm]{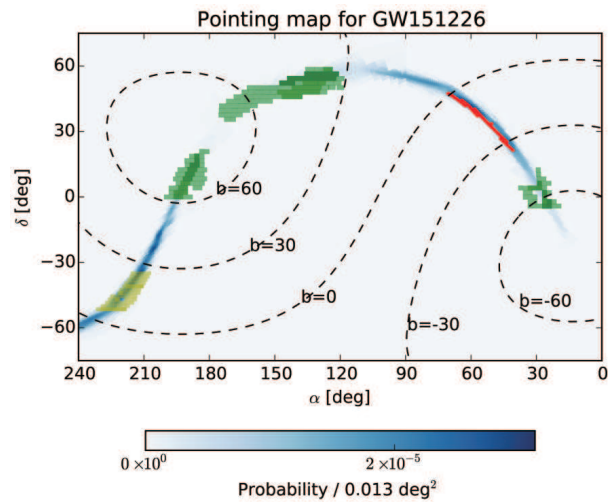} 
 \end{center}
\caption{The observed area of the wide-field surveys of the J-GEM follow-up observation of GW151226
overlaid on the probability skymap (dark blue scale).
Green, red, and yellow colored regions represent the areas observed with KWFC, HSC, and MOA-cam3, respectively.  
}\label{fig:sa}
\end{figure}

The KWFC survey observations were done in $r$-band on Dec. 28 and 29 and
Jan. 1--6 (UT). 
The total area observed with KWFC was 778~deg$^{2}$
far off the Galactic plane. 
To perform an image subtraction with the archival SDSS (Sloan Digital Sky Survey; \cite{Shadab2015}) images, 
the high probability regions
had to be avoided.
Each field was observed typically twice or three times.
The exposure time is 180~sec each and the seeing was
2.5--3.0 arcsec FWHM. 

We carried out an imaging follow-up observations with HSC in the first half nights of 
Jan. 7, 13, and Feb. 6, 2016 (UT). 
We observed an area of 63.5 deg$^2$ centered at ($\alpha$, $\delta$) = (03:33:45, +34:57:14) 
spanning over the highest probability region in the initial skymap (BAYESTAR) with 50 HSC fiducial pointings. 
The fiducial pointings were aligned on a Healpix \citep{Gorski2005} grid with NSIDE=64 
(a corresponding grid size is 0.84 deg$^2$). 
To remove artifacts efficiently, we visited each fiducial pointing twice with a 2 arcmin offset. 
We observed the field in 
$i$-band and 
$z$-band with an exposure time ranging from 45~sec to 60~sec for each pointing. 
On Feb. 6, first we surveyed all the fields by single exposure, then observed the whole
area again. 
The seeing ranged from 0.5 arcsec to 1.5 arcsec FWHM. 

We also performed survey observations with MOA-cam3 for a part of the skymap in 
the southern hemisphere from UT Mar. 8 to 11 2016.
The total area covered by the MOA-cam3 observations was 145 deg$^2$.
The ``MOA-Red'' filter \citep{Sako2008}, which is a special filter dedicated to 
micro-lens survey with a wide range
of transmission from 6200\AA\ to 8100\AA\, was used.
The exposure time per field was 120~sec.
The seeing was 1.9--4.5 arcsec FWHM.

Since the sky areas observed by the three instruments were not overlapped, 
the total area covered by the wide-field surveys was 986.5 deg$^2$. 
The integrated detection probabilities of the observed regions for the final skymap (LALInference)
were 0.07, 0.09, and 0.13 for 
HSC, KWFC, and MOA-cam3,
respectively.
We thus covered a total of $\sim$29\%\ of the probability skymap of GW151226.

\begin{figure}
 \begin{center}
  \includegraphics[width=8cm]{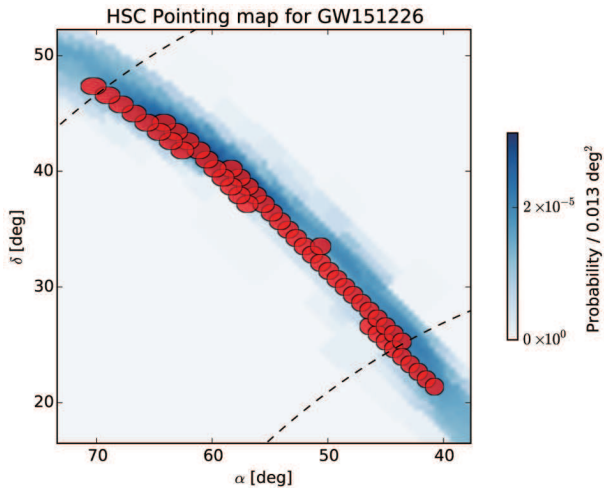} 
 \end{center}
\caption{The enlarged view of the survey area by HSC.}\label{fig:hsa}
\end{figure}

The wide-field survey observations are summarized in Table \ref{tab:survey}.
The survey areas and the probability skymap of GW151226 are shown in Figure \ref{fig:sa}.
An enlarged map of the sky areas observed with HSC is shown in Figure \ref{fig:hsa}.

\begin{table*}
  \tbl{The observing log of the wide field survey observations}{%
  \begin{tabular}{llcccc}
      \hline
      Date (UT) & Instrument &  mid-T$^{\rm a}$ & Area & Band & $m_{\rm lim}$$^{\rm b}$ \\
                    &                 & [days]               & [deg$^2$] &        & [AB mag]           \\ 
      \hline
      2015-12-28 & KWFC    &   2.43 & 176  & $r$       & 19.2$\pm$1.3    \\
      2015-12-29 & KWFC    &   3.48 & 512  & $r$       & 19.5$\pm$0.3    \\
      2016-1-1    & KWFC    &   6.59 & 48   & $r$      & 17.1$\pm$1.2    \\
      2016-1-2    & KWFC    &   7.67 & 124  & $r$       & 20.3$\pm$0.2    \\
      2016-1-3    & KWFC    &   8.70 & 56   & $r$       & 20.1$\pm$0.3    \\
      2016-1-4    & KWFC    &   9.49 & 84   & $r$       & 19.9$\pm$0.3    \\
      2016-1-5    & KWFC    &   10.36 & 40   & $r$       & 19.8$\pm$0.6    \\
      2016-1-6    & KWFC    &   11.60 & 124  & $r$       & 20.0$\pm$0.2    \\
      2016-1-7    & HSC      &   12.71 & 63.5   & $i$, $z$    & $i$: 24.3$\pm$0.2, $z$: 23.5$\pm$0.2  \\
      2016-1-13    & HSC    &  18.17 & 63.5   & $i$, $z$    & $i$: 24.6$\pm$0.2, $z$: 23.8$\pm$0.2   \\
      2016-2-6    & HSC      &  42.17  &  63.5   & $i$, $z$    & $i$: 24.4$\pm$0.2, $z$: 23.8$\pm$0.3   \\
      2016-3-8    & MOA-cam3  &  73.31 &  55  & MOA-red   & 18.2$\pm$0.1  \\
      2016-3-9    & MOA-cam3  &   74.31 & 11  & MOA-red   & 17.3$\pm$1.2   \\
      2016-3-10  & MOA-cam3  &  75.35 & 117  & MOA-red   & 18.2$\pm$0.3   \\
      2016-3-11  & MOA-cam3  &  76.30  & 15  & MOA-red   & 18.2$\pm$0.3   \\
      \hline
    \end{tabular}}\label{tab:survey}
\begin{tabnote}
a. Middle time of the observation in unit of days after GW151226.

b. Median value of 5 $\sigma$ limiting magnitude and its range (1$\sigma$) during one observation run.
\end{tabnote}
\end{table*}

\subsection{Galaxy Targeted Follow-ups}

We performed targeted follow-up imaging observations from UT Dec. 27 2015.
We used seven instruments on six telescopes; HOWPol \citep{Kawabata2008}, 
HONIR \citep{Akitaya2014} on 1.5~m Kanata telescope,
MINT on 2~m Nayuta telescope, MITSuME (MITSuME-OAO) \citep{Kotani2005} on 0.5~m telescope, 
OAO-WFC \citep{Yanagisawa2014} on 0.91~m telescope,
MOA-cam3 on 1.8~m MOA-II telescope and SIRIUS \citep{Nagayama2003} on 1.4~m IRSF, 
for these observations.
We performed $R$ band observations with HOWPol and MITSuME, $I$ band observations
with HOWPol, HONIR, and MINT, MOA-Red observations with MOA-cam3, 
$J$ band observations with OAO-WFC,
and $J$, $H$, and $K$ bands observations with SIRIUS.
 
We selected 309 nearby galaxies from GWGC 
(Gravitational Wave Galaxy Catalog) \citep{White2011} 
in the skymap regions whose detection probabilities are
more than 0.0008. 
We divided the target galaxies into 4 target groups.
The groups 1 to 3 contain northern galaxies accessible from Japan. 
The number of galaxies of the group 1, 2, and 3 are 77, 76, and 77, respectively.
The group 4 contains 79 southern galaxies.
We allocated these groups to the above telescopes as target lists.

\begin{figure}
 \begin{center}
  \includegraphics[width=8cm]{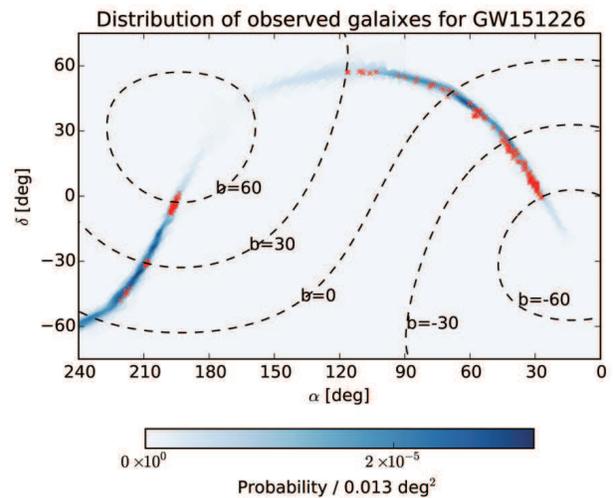} 
 \end{center}
\caption{The positions of the galaxies observed in the J-GEM follow-up observation of 
GW151226 (red points).}\label{fig:gp}
\end{figure}

\begin{figure}
 \begin{center}
  \includegraphics[width=8cm]{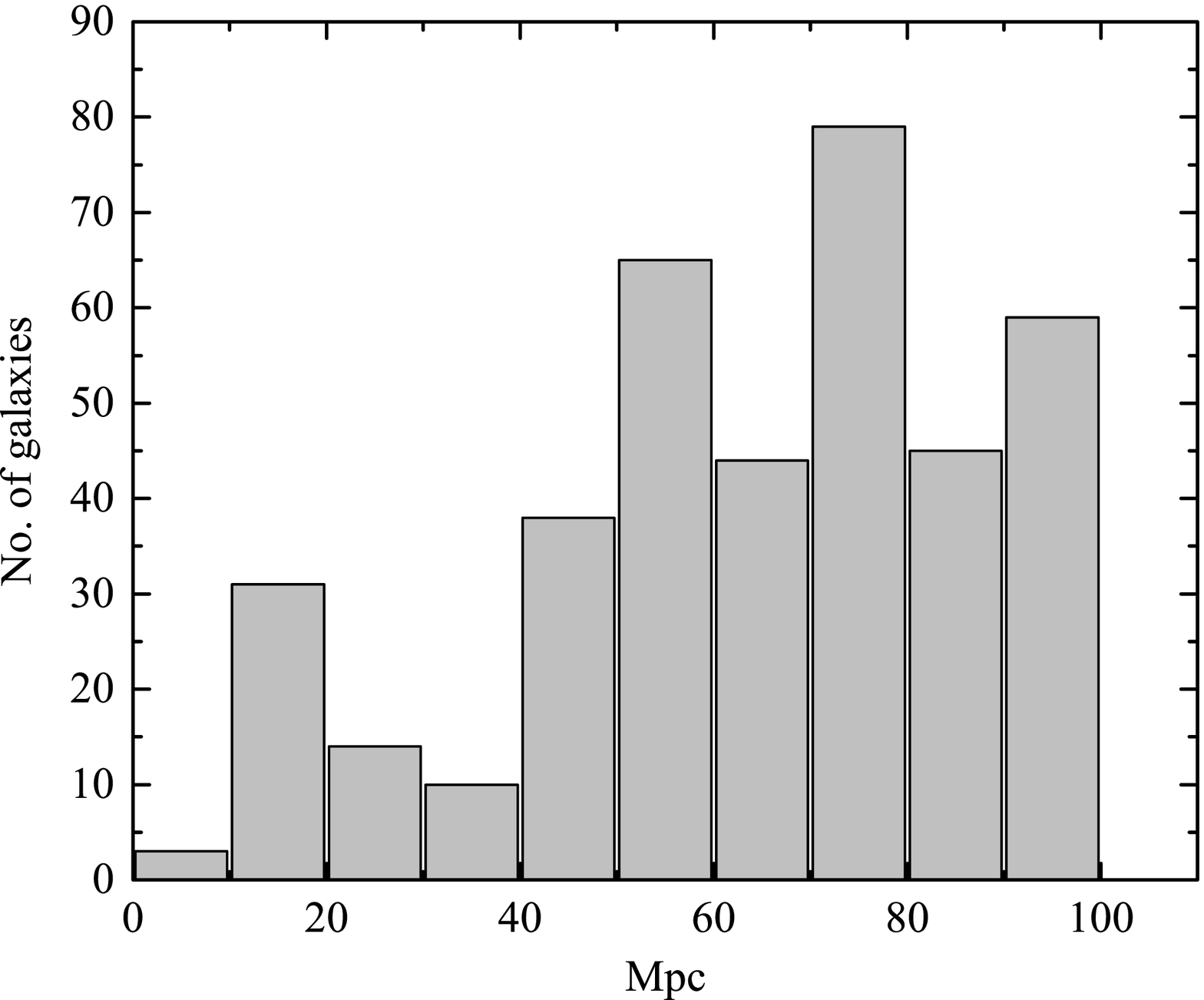} 
 \end{center}
\caption{The distance distribution of the observed galaxies.
The distance limit (100 Mpc)  of the galaxies is determined by GWGC.}\label{fig:gd}
\end{figure}

The summary of the targeted observations is shown in Table \ref{tab:galaxies}.
The net number of the observed galaxies was 238.  
The spatial and distance distributions of the observed galaxies are shown in Figure \ref{fig:gp}
and Figure \ref{fig:gd}, respectively.

\subsection{Spectroscopic Follow-up}

We carried out a spectroscopic observation of MASTER OT J020906.21+013800.1 
\citep{Lipunov2015} with
a fiber-fed integral field spectrograph KOOLS-IFU attached to 
the 188 cm telescope at Okayama Astrophysical Observatory on
UT Dec. 28 2015.
The field of view of KOOLS-IFU is 1.8 arcsec per fiber and 30 arcsec in total.
The wavelength range and spectral resolving power were 5020--8830 \AA\, and
600--850, respectively.
The total exposure time was 3600~sec.

\begin{table*}
  \tbl{The average limiting magnitudes of the galaxy targeted observations}{
  \begin{tabular}{llcccc}
      \hline
      Date (UT) & Instruments & mid-T$^{\rm a}$ & ${N_{\rm gal}}^{\rm b}$ & exp-T & $m_{\rm lim}$$^{\rm c}$ \\ 
                    &                  & [days]              &      & [sec] & [AB] \\   
      \hline
      2015-12-27 & HOWPol    & 1.67 & 18 & 90 & $R$: 17.9$\pm$0.6, $I$: 18.3$\pm$0.4  \\
      2015-12-28 & MITSuME  & 2.46 & 61 & 540 & $R$: 18.5$\pm$0.4     \\
                       & OAO-WFC & 2.46 & 36 & 900 & $J$: 18.3$\pm$0.3 \\
                       & MINT        & 2.47 & 37 &  540 & $I$: 20.1$\pm$0.5  \\
                       & HONIR      & 2.49 & 51 &  120 & $I$: 19.4$\pm$0.5  \\
                       & SIRIUS      & 2.78 & 10 &  360--580  & $J$: 19.3$\pm$0.4, $H$: 19.2$\pm$0.4, $K$: 18.1$\pm$0.4  \\ 
      2015-12-29 & MITSuME  & 3.34 & 16 & 540 & $R$: 18.5$\pm$0.4 \\
                       & MOA-cam3  & 3.45 & 10 & 120 & MOA-red: 17.3$\pm$0.7  \\
                       & OAO-WFC & 3.47 & 32 & 900 & $J$: 16.4$\pm$0.4 \\
                       & HONIR      & 3.49 & 20 & 120 & $I$: 19.7$\pm$0.3  \\
                       & MINT        & 3.53 & 38 & 540 & $I$: 20.0$\pm$0.6  \\
      2015-12-31 & MOA-cam3 & 5.39 & 29 & 120 &  MOA-red: 18.4$\pm$0.1 \\ 
      2016-01-04 & MOA-cam3 & 9.40 & 24 & 120 &  MOA-red: 18.6$\pm$0.2 \\ 
      2016-01-05 & MOA-cam3 & 10.30 & 19 & 120 & MOA-red: 18.2$\pm$0.1 \\ 
      \hline
    \end{tabular}}\label{tab:galaxies}
\begin{tabnote}
a. Middle time of the observation in unit of days after GW151226.

b. Number of observed galaxies.

c. Median value of 5 $\sigma$ limiting magnitude and its range (1$\sigma$) during one observation run.
\end{tabnote}
\end{table*}

\section{Data Reduction and Results}

\subsection{Wide Field Survey Data}

\subsubsection{KWFC survey}

The data reduction of the KWFC data was made using the standard data reduction
pipeline developed for Kiso Supernova Survey (KISS; \cite{Morokuma2014}).
The pipeline functions include bias subtraction, overscan
subtraction, overscan trimming, flat-fielding, point spread function (PSF) size
measurements, astrometry relative to the USNO-B1.0 catalog \citep{Monet2003}, 
zeropoint magnitude determination relative to the SDSS,
image subtraction using the SDSS images,
and detection of transient candidates in the subtracted images.
The 5$\sigma$ limiting magnitudes of the KWFC observations ranged from
18.0 to 20.5 depending on the sky condition of the Kiso observatory.

The transient candidates detected in the subtracted images include not only
astronomical objects but also non-astronomical artifacts,
such as cosmic rays, residual of image subtraction due to
imperfect image alignment or convolution (see e.g. \cite{Bailey2007, Bloom2012}).
Moreover, astronomical objects include minor planets or variable stars in addition to extragalactic transients.
Therefore, we first removed the transient candidates around the objects which are
registered as star in the SDSS catalog.
This effectively removed both variable stars and artifacts around bright stars.
Then, all the sources matched with the database of the Minor Planet Center were removed.
Finally, the remaining objects were visually inspected to remove artifacts.

As a result, we found 13 extragalactic transient candidates associated with galaxies.
The candidates found with the KWFC are summarized in Table \ref{tab:ktrans}.
Nine out of 13 objects are detected more than twice in our survey.
The other four objects (KISS15ah, KISS15ai, KISS16b, and KISS16c) were detected only once.
Since KISS15ah and KISS16c are independently discovered by other groups
(AT 2016bse and SN 2015bl, respectively),
they must be genuine extragalactic transients.
Although there is no independent discovery for KISS15ai and KISS16b,
they are rather bright (16.6 and 19.6 mag, respectively),
and unlikely to be minor planets which are
not registered in the database of the Minor Planet Center.

In Table \ref{tab:ktrans}, we show estimated absolute magnitudes of
13 transient candidates using spectroscopic and photometric redshifts.
Except for KISS16f and KISS16b, the candidates were too bright for the expected kilonova emission
(e.g. \cite{Tanaka2014}), suggesting that they are supernovae (SNe).
KISS16f and KISS16b were rather faint, but their host galaxies
are located at z = 0.012 and 0.009964, respectively,
and thus they were not associated with GW151226.
They are likely to be SNe after the peak brightness.

\begin{table*}
  \tbl{The supernovae identified by the KWFC survey}{%
  \begin{tabular}{lccccccccc}
      \hline
ID & RA & DEC & T$_{\rm obs}$(UT)$^{\rm a}$ & $m_r$ & $m_{\rm lim}$$^{\rm b}$ & host galaxy$^{\rm c}$ & spec-$z$$^{\rm d}$ & photo-$z$ & ${M_r}^{\rm e}$ \\
   & [deg] & [deg] &  & [AB] & [AB] &                   &              &                & [AB] \\
      \hline
KISS15ag & 141.812070 & 51.480666 & 2015-12-28 13:40:48 &17.5  & 19.30  & SDSS J092715.01+512853.2 & 0.053 & - & $-$19.4 \\      
KISS15ah & 140.142947 & 50.696334 & 2015-12-29 12:57:36 & 18.0  & 19.17  & SDSS J092034.44+504148.7 & - & 0.050 or 0.063 & $-$19.1 \\
KISS15ai & 19.249817 & -4.942760 & 2015-12-29 09:50:24 & 16.6  & 20.00  & SDSS J011659.36-045629.0 & - & 0.03 &  $-$19.1 \\
KISS15aj & 137.536390 & 50.061012 & 2015-12-29 12:14:24 & 17.4  & 19.37  & UGC 04812 & 0.0343 & - &  $-$18.6 \\
KISS16a & 126.579910 & 53.770297 & 2016-01-02 18:28:48 & 18.7  & 20.40  & SDSS J082619.18+534610.5 & 0.042 & - &  $-$17.8 \\
KISS16b & 140.725655 & 46.534659 & 2016-01-02 20:52:48 & 19.6  & 20.26  & KUG0919+467  & 0.009964 & - &  $-$13.7 \\
KISS16c & 134.969736 & 53.265282 & 2016-01-02 19:55:12 & 19.3  & 20.46  & SDSS J085952.59+531547.7 & 0.093 & - &  $-$18.9 \\
KISS16d & 136.815119 & 52.762845 & 2016-01-02 19:55:12 & 19.5  & 20.46  & SDSS J090715.76+524544.6 & NA & 0.1 &  $-$18.9 \\
KISS16e & 131.618647 & 53.758743 & 2016-01-02 18:57:36 & 19.8  & 20.18  & SDSS J084628.73+534531.2 & - & 0.10 or 0.08 &  $-$18.4 \\
KISS16f & 140.055455 & 54.108287 & 2016-01-03 20:38:24 & 18.5  & 20.16  & SDSS J092012.28+540628.1 & 0.012 &  &  $-$15.2 \\
KISS16g$^{\rm f}$ & 186.709112 & 16.263777 & 2016-01-03 20:09:36 & 19.7  & 20.41  & SDSS J122649.70+161546.7 & - & 0.55 or 0.26 &  $-$22.1 \\
KISS16h & 126.292102 & 56.706847 & 2016-01-06 19:12:00 & 19.0  & 20.52  & SDSS J082510.12+564222.5 & 0.043 & -  &  $-$17.5 \\
KISS16i & 185.281171 & 16.935903 & 2016-01-06 20:09:36 & 19.7  & 20.41  & SDSS J122107.48+165607.1 & - & 0.1 &  $-$18.7 \\
    \hline
    \end{tabular}}\label{tab:ktrans}
\begin{tabnote}
a. Observation time (UT) of the events.

b. 5 $\sigma$ limiting magnitude.

c. Closest galaxy in SDSS.

d. All the spectral redshifts except for KISS15aj and KISS16b were taken from SDSS DR12 \citep{Shadab2015}.
The redshifts for KISS15aj and KISS16b were obtained from \citet{Fisher1995} and \citet{Falco1999}, respectively.

e. When two values are given for photo-z, an average redshift is assumed.

f. Identification of the host galaxy is uncertain.
  The host galaxy may be SDSS J122650.23+161618.2 ($z=$0.046)
  located at about 29 arcsec north, and
  then the absolute magnitude of the transient is $-$16.1 mag.
\end{tabnote}
\end{table*}

\subsubsection{HSC survey}

The HSC data were reduced using HSC pipeline version 4.0.1, which had been developed
based on the LSST pipeline \citep{Ivezic2008,Axelrod2010}.
The HSC pipeline provides packages for bias subtraction, flat fielding,
astrometry, mosaicing, warping, coadding, and image subtraction.
The astrometry and photometry were made relative to the Pan-STARRS1
(PS1, \cite{Tonry2012}; \cite{Schlafly2012}; \cite{Magnier2013})
with a 1.5 arcsec (9~pixel) aperture diameter.
The limiting magnitudes were estimated by randomly sampling $>10^5$ apertures.

The images taken on Feb. 6 were used as the reference images and were
subtracted from the images taken on Jan. 7 and 13. 
Here, we separately adopted images at 2 epochs on Jan. 7, while images on Jan. 13 were coadded. 
Point sources in the difference images were detected and measured with the HSC pipeline.
Since there were many false detections, 
we screened the detected sources by the following selection procedure.   
(1) In order to exclude false detections, we selected point sources detected in both of
the $z$-band difference images on Jan. 7 at the same location with a signal-to-noise
ratio of $>5\sigma$, ellipticity of $>$0.8, and FWHM of 0.8--1.3 arcsec. 
In addition, a small residual of PSF subtraction from the sources ($<3\sigma$) was imposed. 
(2) To select objects fading from Jan. 7 to Feb. 6, 
fluxes of sources in the two $z$-band difference images on Jan. 7 were required to be positive. 
(3) To exclude minor planets, first we estimated the maximum distance 
that an object could move during an interval between $z$- and $i$-band imaging observations.
We found that it is $\sim$45 arcsec, assuming that the elongation in the $z$-band difference
image was due to the movement of the object during the exposures. 
For the sources survived after the selections (1) and (2), we checked the $i$-band difference images.
If a source was not detected but another transient source was found at a distance of 0.5--45 arcsec
in $i$-band difference image, we omitted the source as a possible minor planet.
We also check the position of sources with MPChecker.

After the above screening, 1256 candidates remained and were visually inspected.
First, we removed clear artifacts from the candidate list by visual inspection.
Then we identified and removed slowly moving objects which are thought to be distant minor
planets not removed by the above criterion (3) by carefully checking the images.
Finally, 60 objects remained as extragalactic transient candidates.

The multicolor light curves of the candidates were derived with forced aperture
photometry of the difference images with 1.5 arcsec aperture diameter.
We corrected the Galactic extinction using \citet{Schlegel1998}.

\begin{figure}
 \begin{center}
  \includegraphics[width=8cm]{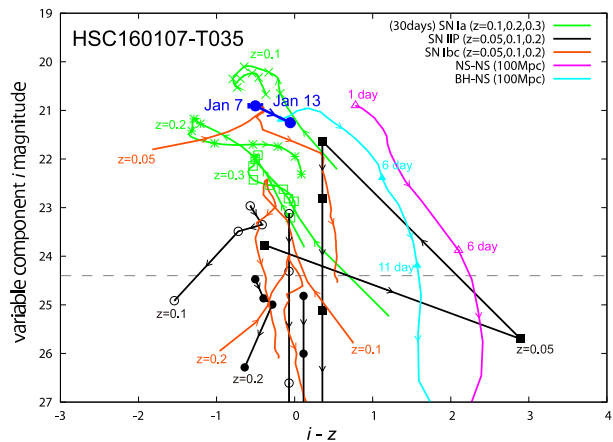} 
 \end{center}
\caption{Color-magnitude variation of variable component of a transient candidate of the 
HSC follow-up survey of GW151226.
Filled blue circles are the data of HSC160107-T035 taken from Jan. 7 and 13 image after
subtracting Feb. 6 image (Galactic extinction was corrected).
Pink and light blue lines represent kilonova models of NS--NS merger and
BH--NS merger of \citet{Tanaka2014} (see text).
Green, black, and orange lines are the color-magnitude evolutions of variable components
of SNe Type Ia, Type IIP, and Type Ibc, respectively.
To derive the variable components of SNe, we subtracted the data 30 days after the
explosions from the model light curves of SNe.
}\label{fig:cm}
\end{figure}

We compared the color-magnitude time variations of variable component of 
the transient candidates between Jan. 7 and 13 with 
the color-magnitude evolutions of Type Ia, Ibc, and IIP SNe and kilonova
emission and classified the candidates (Figure \ref{fig:cm}).
For this comparison, we subtracted the brightnesses at 24 days and 30 days after the explosions
from the model light curves of SNe and simulated the color-magnitude evolutions of
the variable component of SNe.
We adopted fiducial kilonova models of NS--NS merger with ejecta mass of 0.01 $M_\odot$
(model ``APR4-1215'' of \cite{Tanaka2014}) and BH--NS merger with ejecta mass of
0.05 $M_\odot$ (model ``H4Q3a75'' of \cite{Tanaka2014}).

By visual inspection and color-magnitude variation study,
we found that two third of the HSC transients were probably SNe.
One third of the HSC transients were located very close to the centers of the host galaxies
and those time variabilities were not typical of SNe.
We thus classified these sources as ``active galactic nuclei (AGN)''.
No source whose color-magnitude variation is consistent with the kilonova
models was identified by the above procedure.
The extragalactic transient candidates found by the HSC survey
are summarized in Table \ref{tab:htrans}.

\citet{Morokuma2008} derived the number densities of various transient objects
as a function of time interval of $i^\prime$ band observations from
Subaru Suprime-Cam data in the Subaru/XMM-Newton Deep Field \citep{Furusawa2008}.
According to Figure 12 of \citet{Morokuma2008}, the number density of extragalactic transients
(SNe $+$ AGNs) brighter than the variable component $i^\prime$ magnitude $i^\prime_{\rm vari}$ of 25 mag
with 30 days interval observations is $\sim$30. 
The variable component $i$ band limiting magnitude and the number density
of the extragalactic transients in our work are $\sim$24 mag (see Table \ref{tab:htrans})
and $\sim$1 deg$^{-2}$, respectively.
Scaling the number density of \citet{Morokuma2008} using Figure 13 of
\citep{Morokuma2008}, we estimate that it would be 3--4 deg$^{-2}$ for 
the limiting magnitude $i^\prime_{\rm vari} \sim 24$ mag.
This is a few times higher than the value of our observation.
Part of this discrepancy would come from our detection strategy.
We detected the transients based on $z$-band observation, thus we could systematically undercount
blue transients.
In addition,  since the Galactic latitude $b$ of the HSC observation field is less than $\sim$30 deg.
(see Figure \ref{fig:sa}),
large fraction of the field suffered from Galactic extinction
(typical color excess $E$($B-V$) is $\sim$0.3--0.7; see Table \ref{tab:htrans}).
Considering these factors, we judge that our observation is roughly consistent with 
\citet{Morokuma2008}.

\begin{longtable}{lcccccccc}
  \caption{The extragalactic transients identified by the HSC survey}\label{tab:htrans}
  \hline
  ID &   RA    & DEC & $E$($B-V$) & $T_{\rm obs}$($i$)$^{\rm a}$ & $m_i$  & $T_{\rm obs}$($z$)$^{\rm b}$ & $m_z$ & type \\
      &   [deg] & [deg] &           &                      & [AB] &                       & [AB] &      \\
\endfirsthead
  \hline
  ID &   RA    & DEC & $E$($B-V$) & $T_{\rm obs}$($i$) & $m_i$  & $T_{\rm obs}$($z$) & $m_z$ & type \\
      &   [deg] & [deg] &           &       & [AB] &           & [AB] &      \\
  \hline
\endhead
  \hline
\endfoot
  \hline
\multicolumn{9}{l}{\small a. Observation time (UT) in $i$ band on Jan. 7 2016.}\\
\multicolumn{9}{l}{\small b. Observation time (UT) in $z$ band on Jan. 7 2016.}\\
\endlastfoot
  \hline
HSC160107-T001 & 40.997379  & 22.369333  & 0.21 & 09:22:40 & 23.9    & 07:33:46 & 22.5 & SN \\
HSC160107-T002 & 41.176235  & 22.611018  & 0.27 & 09:22:40 & 22.9    & 07:33:46 & 22.7 & SN \\
HSC160107-T003 & 42.560811  & 23.350175  & 0.21 & 09:18:45 & 24.4    & 07:29:08 & 24.0 & SN \\
HSC160107-T004 & 42.872344  & 22.315740  & 0.41 & 09:20:04 & $>$24.2 & 07:30:41 & 22.2 & SN \\
HSC160107-T005 & 43.455010  & 25.258338  & 0.12 & 09:12:10 & $>$24.5 & 07:21:26 & 22.9 & AGN \\
HSC160107-T006 & 43.507674  & 24.850162  & 0.12 & 09:11:31 & $>$24.7 & 07:20:39 & 22.8 & SN \\
HSC160107-T007 & 43.754581  & 23.637964  & 0.23 & 09:14:50 & 23.7    & 07:24:31 & 23.1 & SN \\
HSC160107-T008 & 44.116261  & 24.054421  & 0.14 & 09:12:12 & 21.7    & 07:21:26 & 21.2 & SN \\
HSC160107-T009 & 44.136838  & 25.945316  & 0.12 & 09:06:57 & $>$24.0 & 07:15:12 & 21.7 & SN \\
HSC160107-T010 & 44.364382  & 24.190641  & 0.13 & 09:11:32 & 24.0    & 07:20:39 & 22.8 & AGN  \\
HSC160107-T011 & 44.752975  & 26.107955  & 0.21 & 09:05:38 & 23.4    & 06:21:47 & 23.1 & SN \\
HSC160107-T012 & 44.819914  & 24.395057  & 0.22 & 09:09:33 & $>$23.9 & 07:18:20 & 22.2 & AGN \\
HSC160107-T013 & 45.332537  & 25.263094  & 0.31 & 08:59:07 & $>$24.7 & 05:21:51 & 23.0 & AGN \\
HSC160107-T014 & 45.382080  & 24.835433  & 0.27 & 09:01:43 & 24.5    & 05:25:14 & 23.1 & SN \\
HSC160107-T015 & 45.692939  & 26.530651  & 0.19 & 08:57:51 & 24.1    & 05:18:53 & 24.1 & SN  \\
HSC160107-T016 & 45.985724  & 27.425493  & 0.19 & 08:56:33 & $>$25.0 & 05:18:28 & 22.3 & AGN  \\
HSC160107-T017 & 46.008330  & 25.975611  & 0.22 & 08:53:58 & 23.6    & 05:15:23 & 23.3 & SN \\
HSC160107-T018 & 46.099802  & 27.108579  & 0.19 & 08:55:16 & 24.3    & 05:16:55 & 22.5 & AGN  \\
HSC160107-T019 & 46.346789  & 26.882343  & 0.21 & 08:55:16 & 22.5    & 05:16:55 & 22.4 & SN \\
HSC160107-T020 & 46.462762  & 27.009164  & 0.21 & 08:56:20 & 21.6    & 05:18:12 & 21.2 & SN \\
HSC160107-T021 & 46.830698  & 27.322635  & 0.21 & 08:53:58 & $>$24.3 & 05:15:22 & 22.8 & SN \\
HSC160107-T022 & 47.162617  & 28.111701  & 0.29 & 08:51:23 & 23.5    & 05:12:15 & 23.1 & SN \\
HSC160107-T023 & 47.180281  & 28.363844  & 0.25 & 08:48:49 & 24.2    & 05:09:08 & 23.9 & SN \\
HSC160107-T024 & 47.648348  & 28.246272  & 0.49 & 08:48:49 & 24.0    & 05:09:08 & 22.6 & SN \\
HSC160107-T025 & 47.734609  & 28.924534  & 0.37 & 08:47:30 & 23.7    & 05:07:35 & 22.8 & SN \\
HSC160107-T026 & 47.762266  & 29.189132  & 0.29 & 08:46:51 & $>$24.3 & 05:06:49 & 22.7 & SN \\
HSC160107-T027 & 48.584401  & 30.219543  & 0.37 & 08:41:28 & $>$24.8 & 05:45:30 & 23.6 & AGN \\
HSC160107-T028 & 48.878845  & 30.786932  & 0.37 & 08:37:52 & 21.7    & 07:11:24 & 21.5 & SN \\
HSC160107-T029 & 50.365169  & 33.849423  & 0.24 & 08:40:38 & 21.0    & 04:59:49 & 20.8 & SN \\
HSC160107-T030 & 50.453222  & 32.469045  & 0.41 & 08:28:22 & $>$24.3 & 07:00:39 & 23.3 & SN \\
HSC160107-T031 & 50.621347  & 32.624719  & 0.38 & 08:27:00 & 22.7    & 06:59:05 & 22.2 & AGN  \\
HSC160107-T032 & 50.830253  & 32.696495  & 0.40 & 08:25:38 & $>$24.6 & 06:57:32 & 22.6 & AGN  \\
HSC160107-T033 & 50.892772  & 32.243608  & 0.38 & 08:27:00 & 23.1    & 06:59:05 & 23.1 & SN \\
HSC160107-T034 & 51.672064  & 33.625310  & 0.27 & 08:18:53 & 23.0    & 06:50:19 & 22.6 & AGN  \\
HSC160107-T035 & 52.595560  & 35.179117  & 0.29 & 08:10:29 & 21.4    & 06:41:00 & 21.7 & SN \\
HSC160107-T036 & 53.315983  & 35.731965  & 0.27 & 08:11:54 & $>$24.7 & 06:42:33 & 23.6 & SN \\
HSC160107-T037 & 53.909867  & 35.092927  & 0.34 & 08:11:54 & 23.9    & 06:42:33 & 23.5 & SN \\
HSC160107-T038 & 54.092770  & 35.448804  & 0.30 & 08:11:54 & $>$24.3 & 06:42:33 & 22.2 & SN \\
HSC160107-T039 & 54.585872  & 37.015130  & 0.52 & 08:22:12 & $>$24.1 & 06:53:56 & 22.4 & SN \\
HSC160107-T040 & 54.912712  & 36.394118  & 0.44 & 08:22:12 & $>$23.8 & 06:53:56 & 20.9 & AGN  \\
HSC160107-T041 & 55.370525  & 37.555876  & 0.44 & 08:52:46 & $>$24.6 & 06:36:00 & 23.3 & AGN  \\
HSC160107-T042 & 55.632338  & 36.242112  & 0.49 & 08:22:12 & $>$24.4 & 06:53:56 & 22.9 & SN \\
HSC160107-T043 & 56.537885  & 38.800077  & 0.32 & 09:31:06 & 23.2    & 05:35:10 & 24.0 & AGN  \\
HSC160107-T044 & 56.639089  & 36.644814  & 0.40 & 09:33:41 & 23.2    & 05:38:16 & 22.8 & SN \\
HSC160107-T045 & 56.898156  & 36.857295  & 0.36 & 09:33:41 & 22.1    & 05:38:16 & 22.7 & AGN  \\
HSC160107-T046 & 57.003385  & 36.936598  & 0.34 & 09:33:41 & 22.5    & 05:38:16 & 22.6 & SN \\
HSC160107-T047 & 57.024877  & 36.695131  & 0.38 & 09:33:41 & 21.2    & 05:38:16 & 21.9 & SN \\
HSC160107-T048 & 58.172853  & 37.840891  & 0.95 & 09:38:58 & 23.5    & 05:44:25 & 22.8 & SN \\
HSC160107-T049 & 60.477172  & 39.860675  & 0.83 & 09:49:23 & 22.3    & 05:56:43 & 21.6 & AGN  \\
HSC160107-T050 & 62.176935  & 42.152778  & 0.58 & 09:58:26 & 23.4    & 06:07:32 & 23.6 & SN \\
HSC160107-T051 & 63.477258  & 41.424544  & 0.73 & 09:59:44 & 23.2    & 06:09:05 & $>$23.2 & SN \\
HSC160107-T052 & 64.308645  & 42.773320  & 0.78 & 10:07:36 & 24.7    & 06:18:34 & 22.5 & AGN  \\
HSC160107-T053 & 64.875372  & 43.850244  & 0.79 & 10:10:14 & 25.1    & 06:21:41 & 23.0 & AGN  \\
HSC160107-T054 & 65.638499  & 43.614708  & 0.70 & 10:12:52 & 22.3    & 06:24:49 & 22.4 & SN \\
HSC160107-T055 & 66.332247  & 44.279330  & 0.80 & 10:14:11 & 22.2    & 06:26:22 & 21.5 & SN \\
HSC160107-T056 & 67.121767  & 45.254756  & 1.48 & 10:17:29 & $>$24.3 & 06:29:26 & 20.7 & AGN  \\
HSC160107-T057 & 67.213427  & 45.250006  & 1.52 & 10:16:50 & 23.4    & 06:29:26 & 22.5 & SN \\
HSC160107-T058 & 69.108532  & 46.036008  & 1.78 & 10:20:46 & 22.8    & 06:34:05 & 21.8 & SN \\
HSC160107-T059 & 69.776861  & 46.009513  & 1.55 & 10:20:46 & $>$24.5 & 06:34:05 & 22.2 & SN \\
HSC160107-T060 & 69.983965  & 47.715348  & 1.40 & 10:23:25 & 22.8    & 06:37:13 & 21.2 & AGN  \\
\hline

\end{longtable}

\subsubsection{MOA-II survey}

The data of MOA-II were reduced in standard manner of CCD data reduction using IRAF.
Astrometry of the data was done using Astrometry.net \citep{Lang2009}.
Then point source candidates were extracted with SExtractor \citep{Bertin1996}. 
After excluding known stars using the USNO-B1.0 catalog,
we omitted the candidates whose brightness profiles were not consistent with
PSF by profile fitting using IRAF task ALLSTAR.
We visually inspected the remaining 2953 candidates and selected 39
sources as transient object candidates.
Then we checked 2MASS \citep{Skrutskie2006} and WISE \citep{Wright2010} images
and found that 33 among the 39 candidates were 2MASS sources and one was a WISE source. 
Using Minor Planet Checker (MPChecker)\footnote{http://www.minorplanetcenter.net/cgi-bin/checkmp.cgi}, 
we found 3 candidates were asteroids. 
One of the candidates was a known supernova PSN J14102342-4318437.

After all selections, one candidate with $\sim$18.0 mag located at ($\alpha$, $\delta$)
$=$ (14:44:41.06, $-$44:4:38.4) remained.
This source did not seem to be associated with bright galaxies. 
We observed this source twice with the interval of 180~sec on Mar. 10 2016 and
did not detect significant motion of it between the two exposures. 
It completely disappeared at the third observation performed at the end of Aug. 2016. 
Though we cannot exclude the possibility that this source is an extragalactic transient, 
we think that the most plausible explanation is a minor planet not cataloged in MPChecker.

In the above processing, faint objects embedded in galaxies could be systematically lost.
To detect such sources, we selected 2143 galaxies between 250 and 620 Mpc in the observed fields
using GLADE
(Galaxy List for the Advanced Detector Era)\footnote{http://aquarius.elte.hu/glade}. 
We found 549 point sources within 5 arcsec around these galaxies. 
Compared to DSS images, we found all the sources were known objects.

\subsection{Galaxy Targeted Follow-up Data}

The data reduction of the instruments used for the galaxy targeted observations ---
HOWPol, HONIR, MINT, MITSuME, MOA-II, OAO-WFC, and SIRIUS --- was
made in a standard manner; overscan correction, bias and dark subtraction, and
flat-fielding.
Then multiple exposure frames were coadded. 
The photometric calibrations of the optical data were made by comparing the fluxes of 
the field stars with those listed in the SDSS or 
GSC2.3 (Guide Star Catalog version 2.3).
For the near-infrared bands data calibration, we used the 2MASS point source catalog (PSC)
\citep{Skrutskie2006}.
The observed galaxies and the limiting magnitudes of our observations are listed
in Figure 1 of on-line supplementary data.

We searched for transient point sources in the observed frames taken 
with the above instruments by comparing them
with DSS red frames for $R$ and $I$ bands, and with 2MASS PSC for near-infrared bands.
We found transient candidates in $I$ band frames of the galaxies
PGC1202981 and UGC~1410 taken with HONIR on Dec. 28 2015.
However, the former one was a Galactic variable star and the latter was a known minor planet.
We also found a possible transient candidate close to the nucleus of PGC1021744 in
a $J$ band image taken
with OAO-WFC on Dec. 28 2015.
Since the source was slightly fainter than the 5$\sigma$ limiting magnitude of
the image ($\sim$17.2 mag), the detection was quite marginal.
We made a follow-up observation for this object with OAO-WFC in the next night.
The limiting magnitude of the observation reached 19.2 mag in $J$ band with an exposure
time of 2700~sec, but no point source was found at the same position.
We thus could not confirm whether the source was a real astronomical transient.

As a conclusion, no extragalactic transient object was found with our
galaxy targeted follow-up of GW151226.

\subsection{Spectroscopic Follow-up Data}

The target of the spectroscopy, MASTER OT J020906.21+013800.1, 
was reported at the unfiltered magnitude of 18.3 in the skymap area of GW151226
on UT Dec. 27 2015 and
reported to be brightening \citep{Lipunov2015}.
Our integral field spectroscopy found no significant signal from the OT candidate.
Given that the radial intensity profile of the object is the Gaussian with FWHM of 3 arcsec,
$\sim$40\% of the object flux falls in 3 fibers.
The 5$\sigma$ limiting magnitude was 17.4 at 7400 \AA.
It is noted that the observations with the 3.6 m TNG starting on UT Dec. 28.8247 2015
also did not find any evidence for the OT with the upper limit of $r = 21.0$ mag \citep{DAvanzo2015}.
They detected the emission from a faint galaxy at the redshift of $\sim$0.034
at the position of the OT \citep{DAvanzo2015}.


\section{Discussion and Conclusion}

No optical and near-infrared counterpart of the gravitational wave event GW151226 
was identified by the
follow-up observations under the J-GEM collaboration.
Other teams' trials to find EM counterparts associated with this event also failed
\citep{Adriani2016,Cowperthwaite2016,Evans2016b,Racusin2016,Smartt2016b}.
We found 13 SNe candidates in the KWFC survey data, and 60 extragalactic transients
in the HSC survey data.
About two third of the HSC transients were probably SNe, and the remaining one third 
were classified as possible AGNs.
There was no source which showed the color-magnitude variation consistent with
current kilonova models in our dataset.
We thus conclude that this work did not find clear candidates of EM
counterpart of the gravitational wave source.

Both of the two GW events, GW150914 and GW151226, detected by aLIGO were BH--BH mergers.
Inspired by the possible detection of a $\gamma$-ray emission associated with GW150914
by $Fermi$ satellite \citep{Connaughton2016}, several physical mechanisms 
for EM emission from a BH--BH merger event have been proposed \citep{Morsony2016,Perna2016,Yamazaki2016}.
However, all of those theoretical works have difficulties in producing strong EM emission by a
BH--BH merger.
In addition, questions have been raised for the reality of the $\gamma$-ray detection by $Fermi$
both from theoretical side \citep{Lyutikov2016,Zhang2016} and observational and data analysis side 
\citep{Greiner2016,Savchenko2016,Xiong2016}.
Thus there is still no observational evidence with 
concrete theoretical background for EM emission from
BH--BH merger.
In other words, the key ingredient for detection of EM counterpart associated with GW is 
whether it contains a neutron star.
Hence the information of the chirp mass of GW event is crucial for EM follow-up observations.
When the chirp mass and distance estimation of a GW event is distributed, 
EM follow-up teams will be able to make effective observation plans with their available
facilities \citep{Singer2016}.

\begin{figure}
 \begin{center}
  \includegraphics[width=8cm]{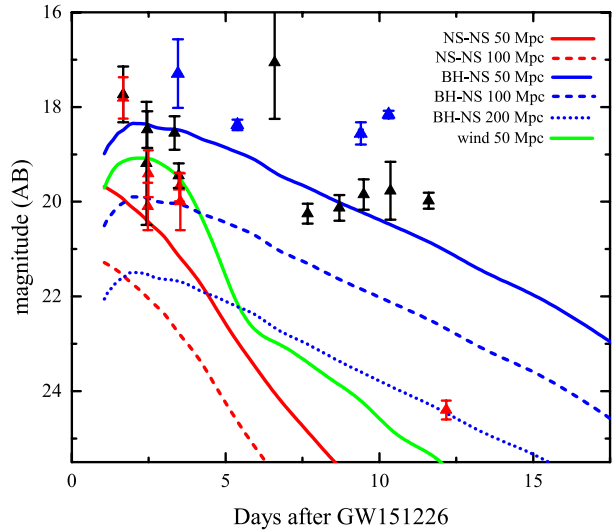} 
 \end{center}
\caption{The limiting magnitudes of the J-GEM observations of GW151226 and
kilonova light curves.
Filled triangles represent median 5$\sigma$ limiting magnitudes and the y-axis error bars
show the range of the variation of the limiting magnitudes in the observed data sets.
Black, red, and blue colors represent $R$ ($r$ for KWFC) band, $I$ ($i$ for HSC) band, 
and MOA-red band, respectively.
The theoretical $i$ band light curves of NS--NS merger (APR4-1215 of \cite{Tanaka2014})
and BH--NS merger (H4Q3a75 of \cite{Tanaka2014}) are shown as red and blue lines, respectively.
The green line shows the $i$ band light curve of a model of the emission from
shocked wind from NS--NS merger with ejecta mass of 0.03 $M_\odot$ \citep{Tanaka2016}.
Solid, dashed and dotted lines correspond to the event distance of 50 Mpc, 100 Mpc and 200 Mpc,
respectively.}\label{fig:obslc}
\end{figure}

For considering future observation strategies, we summarized the observation epochs and
the limiting magnitudes of the J-GEM follow-up of GW151226 in Figure \ref{fig:obslc}.
The limiting magnitudes of $R$, $r$, $I$, $i$, and MOA-red bands taken with 
HOWPol, HONIR, MINT, MITSuME, MOA-cam3, KWFC, and HSC are plotted 
with theoretical $i$ band light curves of kilonovae \citep{Tanaka2014,Tanaka2016}.
Our early observations with the small- and mid-sized- telescopes reached the depth of $\sim$20 mag
in optical red bands.
The KWFC data around 6--8 days after the GW event were as deep as $\sim$20.5 mag.
The deepest data taken with HSC reached down to $\sim$24 mag in $i$ band at 12 days after
GW151226.
According to the theoretical light curves in Figure \ref{fig:obslc}, the depth of our early galaxy targeted 
observations reached the detection threshold of kilonova emission from a BH--NS merger
within a distance of $\sim$50--100 Mpc.
The late KWFC observations at around 7 days after the GW could follow the candidate.
The deep HSC observations could follow the light curve of the candidate at most one month
after the event.

However, if the event were NS--NS merger, the story would be completely changed.
The kilonova emission for NS--NS merger is too faint to detect with our observations.
Even if the event distance is 50 Mpc, the maximum magnitude of the optical emission
would be much fainter than $\sim$19 mag at 1 day after the event.
Only HSC could detect the optical emission from a kilonova at a distance of 50--100 Mpc
if the follow-up observation with HSC was performed within $\sim$5 days after the event.

HSC has a capability to survey over $\sim$60 deg$^2$ with two colors, $i$ and $z$ bands, with 
the limiting magnitude of $\sim$24 mag within a half night.
Figure \ref{fig:obslc} shows that quick ($<$3 days) follow-up observations with HSC can detect 
the optical emission of kilonova 
induced by an NS--NS merger at a distance of $\sim$200 Mpc.
For BH--NS mergers, relatively slow start of the observation is acceptable.
The kilonova EM emission from BH--NS merger at a distance of 400 Mpc would be 
detectable by HSC even after 10 days from the GW event.
When aVirgo goes in regular operation and a joint observation of aLIGO and aVirgo starts,
the 90\%\ credible area of GW detection would become smaller than $\sim$50 deg$^2$
depending on the signal-to-noise ratio of the event \citep{Singer2014}.
This size of area matches very well to the area covered by half night observation of HSC,
and thus detection of EM emission from kilonova is greatly anticipated.

\begin{ack}
This work makes use of software developed for the Large Synoptic Survey Telescope. 
We thank the LSST Project for making their code available as free software at http://dm.lsstcorp.org.
The Pan-STARRS1 Surveys (PS1) have been made possible through contributions of the Institute for Astronomy, the University of Hawaii, the Pan-STARRS Project Office, the Max-Planck Society and its participating institutes, the Max Planck Institute for Astronomy, Heidelberg and the Max Planck Institute for Extraterrestrial Physics, Garching, The Johns Hopkins University, Durham University, the University of Edinburgh, Queen's University Belfast, the Harvard-Smithsonian Center for Astrophysics, the Las Cumbres Observatory Global Telescope Network Incorporated, the National Central University of Taiwan, the Space Telescope Science Institute, the National Aeronautics and Space Administration under Grant No. NNX08AR22G issued through the Planetary Science Division of the NASA Science Mission Directorate, the National Science Foundation under Grant No. AST-1238877, the University of Maryland, and Eotvos Lorand University (ELTE).
This research has made use of the NASA/IPAC Extragalactic Database (NED) which is operated by the Jet Propulsion Laboratory, California Institute of Technology, under contract with the National Aeronautics and Space Administration.
This work was supported by MEXT Grant-in-Aid for Scientific Research on Innovative Areas 
"New Developments in Astrophysics Through Multi-Messenger Observations of Gravitational Wave Sources"
(JP24103003), JSPS KAKENHI Grant Numbers JP26800103 and JP15H02069,
and the research grant program of Toyota foundation (D11-R-0830). 
\end{ack}


\end{document}